\documentclass[preprint,aps]{revtex4}

\newcommand {\bea}{\begin{eqnarray}}
\newcommand {\eea}{\end{eqnarray}}
\newcommand {\be}{\begin{equation}}
\newcommand {\ee}{\end{equation}}

\usepackage{graphicx}

\begin{document}


\title{Non-Fermi Liquid Effects in QCD at High Density}

\author{Thomas~Sch\"afer$^{1,2}$ and Kai~Schwenzer$^1$}

\affiliation{
$^1$Department of Physics, North Carolina State University,
Raleigh, NC 27695\\
$^2$Riken-BNL Research Center, Brookhaven National 
Laboratory, Upton, NY 11973}

\begin{abstract}
 We study non-Fermi liquid effects due to the exchange of 
unscreened magnetic gluons in the normal phase of high density 
QCD by using an effective field theory. A one-loop calculation
gives the well known result that magnetic gluons lead
to a logarithmic enhancement in the fermion self energy 
near the Fermi surface. The self energy is of the 
form $\Sigma(\omega)\sim \omega\gamma\log(\omega)$, where 
$\omega$ is the energy of the fermion, $\gamma=O(g^2)$,
and $g$ is the coupling constant. Using an analysis of 
the Dyson-Schwinger equations we show that, in the weak 
coupling limit, this result is not modified by higher order 
corrections even in the regime where the logarithm is 
large, $\gamma\log(\omega)\sim 1$. We also show that 
this result is consistent with the renormalization group
equation in the high density effective field theory.

\end{abstract}
\maketitle

\newpage

\section{Introduction}
\label{sec_intro}

  It is theoretically well established that dense quark matter is 
not a Fermi liquid. Attractive interaction between pairs of quarks 
that are anti-symmetric in color cause an instability in the 
quark-quark scattering amplitude if the quark momenta lie on opposite 
sides of the Fermi surface \cite{Bailin:1984bm,Alford:1998zt,Rapp:1998zu}.
This instability is resolved by the formation of a diquark 
condensate which breaks the color gauge symmetry. It is also well 
known that this is not the only non-Fermi liquid effect in dense 
quark matter. Unscreened magnetic gluon exchanges lead to a logarithmic 
singularity in the quark self-energy close to the Fermi surface 
\cite{Baym:uj,Vanderheyden:1996bw,Manuel:2000mk,Manuel:2000nh,Brown:2000eh,Boyanovsky:2000bc,Boyanovsky:2000zj,Ipp:2003cj}.
This logarithmic singularity may lead to a breakdown of 
perturbation theory in the normal phase of dense QCD at 
very low energies, $\omega\sim \mu\exp(-c_{\it nfl}/g^2)$, where 
$\mu$ is the chemical potential, $g$ is the coupling constant, 
and $c_{\it nfl}=9\pi^2$.

 This scale is exponentially small as compared to the scale 
of superconductivity, $\omega\sim \mu\exp(-c_{bcs}/g)$, where $c_{bcs}
=3\pi^2/\sqrt{2}$ \cite{Son:1998uk}. Understanding non-Fermi liquid
effects in the normal phase of quark matter is nevertheless
important. First of all, understanding the normal phase is 
necessary in order to put calculations in the superfluid phase
on a solid footing. Also, in order to establish the possible
existence of a superconducting phase of quark matter from the 
observation of neutron stars we have to compute the properties
of both the normal and the superconducting phase. And finally, 
non-Fermi liquid effects may play a role if the dominant 
superconducting phase is suppressed by non-zero quark masses, 
lepton chemical potentials, or magnetic fields. 

 If electric charge neutrality is taken into account, then a non-zero 
strange quark mass leads to approximately equal differences
between the Fermi momenta of strange and up as well as up and 
down quarks \cite{Alford:2002kj}. This implies that if the strange 
quark mass exceeds a critical value, only pairing between quarks
of the same flavor is possible. Pairing between equal flavors
requires order parameters with non-zero spin, and the 
corresponding gaps are suppressed by roughly two orders
of magnitude compared to the spin zero gap \cite{Schafer:2000tw}.
The gap can be further suppressed by a non-zero temperature
or magnetic field. Finally, flavor symmetry breaking may 
lead to the appearance of gapless fermion modes even in the 
superconducting phase \cite{Shovkovy:2003uu,Alford:2003fq}.
Depending on whether there is magnetic screening in this 
phase the gapless modes will also lead to interesting 
non-Fermi liquid effects.

 In this work we study non-Fermi liquid effects in QCD using the 
high density effective theory 
\cite{Hong:2000tn,Hong:2000ru,Nardulli:2002ma,Schafer:2003jn}.
The paper is organized as follows. In Sect.~\ref{sec_hdet}
we discuss power counting in the high density effective 
theory. In Sects.~\ref{sec_ds} and \ref{sec_num} we study the 
Dyson-Schwinger equation for the quark self energy in the normal 
phase of dense quark matter. In Sect.~\ref{sec_rg} we consider the 
renormalization group equation for the quark propagator.
We discuss some of the implications of our results in 
Sect.~\ref{sec_sum}. Non-Fermi liquid effects due to unscreened 
gauge boson exchanges were first discussed by Holstein, Norton 
and Pincus in the case of a cold electron gas 
\cite{Holstein:1973,Reizer:1989,Gan:1993,Chakravarty:1995}. Similar 
effects due to dynamical gauge fields in systems of strongly correlated
electrons were studied by Polchinski \cite{Polchinski:ii},
Nayak and Wilczek \cite{Nayak:ng}, and others. 

\section{High Density Effective Theory}
\label{sec_hdet}

  At high baryon density the relevant degrees of freedom are 
particle and hole excitations which move with the Fermi 
velocity $v$. Since the momentum $p\sim v\mu$ is large, 
typical soft scatterings cannot change the momentum by very 
much and the velocity is approximately conserved. An effective 
field theory of particles and holes in QCD is given by 
\cite{Hong:2000tn,Hong:2000ru}
\be
\label{l_hdet}
{\cal L} =\sum_{v}
 \psi_{v}^\dagger (iv\cdot D) \psi_{v} 
 -\frac{1}{4}G^a_{\mu\nu} G^a_{\mu\nu}+ \ldots ,
\ee
where $v_\mu=(1,\vec{v})$. The field $\psi_v$ describes particles 
and holes with momenta $p=\mu\vec{v}+l$, where $l\ll\mu$. We will 
write $l=l_0+l_{\|}+l_\perp$ with $\vec{l}_{\|}=\vec{v}(\vec{l}
\cdot \vec{v})$ and $\vec{l}_\perp = \vec{l}-\vec{l}_{\|}$. 
In order to take into account all low energy degrees of freedom
we have to cover the Fermi surface with patches labeled by the 
local Fermi velocity. 

 Higher order terms in the effective lagrangian are suppressed 
by inverse powers of the chemical potential. There are two 
types of higher order corrections, operators that only involve
fields in a given patch, and operators with four or more 
fermion fields that connect fields in different patches
\cite{Schafer:2003jn}. In order to understand the importance
of higher order corrections we have to develop a power counting 
scheme for the high density effective theory. We first discuss a 
``naive'' attempt to count powers of the small scale $l$. In 
the naive power counting we assume that $v\cdot D$ scales as 
$l$, $\psi_v$ scales as $l^{3/2}$, $A_\mu$ scales as $l$, and 
every loop integral scales as $l^4$. We also assume that 
$\vec{D}_\perp,\bar{v}\cdot D\sim l$, where $\bar{v}_\mu=(1,
-\vec{v})$. In this case it is easy to see that a general 
diagram with $V_k$ vertices of scaling dimension $k$ scales 
as $l^\delta$ with
\be
\label{pc_naive}
\delta = 4 +\sum_k V_k(k-4).
\ee
A general vertex is of the form
\be
\label{dim_count}
\psi^a(v\cdot D)^b(\bar{v}\cdot D)^c (D_\perp)^d
 (1/\mu)^e,
\ee
and has mass dimension $3a/2+b+c+d-e=4$. Since $k=3a/2+b+c+d$
and $e\geq 0$ we have $k-4\geq 0$. This implies
that the power counting is trivial: All diagrams constructed
from the leading order lagrangian have the same scaling, 
all diagrams with higher order vertices are suppressed,
and the degree of suppression is simply determined by 
the number and the scaling dimension of the vertices.

\begin{figure}
\includegraphics[width=6cm]{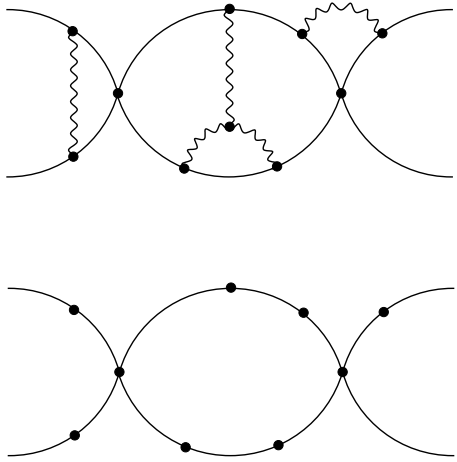}
\caption{Counting hard loops in the effective field theory.
If all (soft) gluon lines are removed the remaining fermionic
loops contain sums over the velocity index.}
\label{fig_lcount}
\end{figure}

 Complications arise because not all loop diagrams
scale as $l^4$. In fermion loops sums over patches 
and integrals over transverse momenta can combine 
to give integrals that are proportional to the 
surface area of the Fermi sphere, 
\be
\label{hard_int}
 \frac{1}{2\pi}\sum_{v}\int\frac{d^2l_\perp}{(2\pi)^2}
 =\frac{\mu^2}{2\pi^2}\int \frac{d\Omega}{4\pi}.
\ee
These loop integrals scale as $l^2$, not $l^4$. In the 
following we will refer to loops that scale as $l^2$
as ``hard loops'' and loops that scale as $l^4$ as ``soft 
loops''. In order to take this distinction into account we 
define $V_k^S$ and $V_k^H$ to be the number of soft and hard 
vertices of scaling dimension $k$. A vertex is called soft if it 
contains no fermion lines. In order to determine the $l$ 
counting of a general diagram in the effective theory we 
remove all gluon lines from the graph, see Fig.~\ref{fig_lcount}. 
We denote the number of connected pieces of the remaining graph 
by $N_C$. Using Euler identities for both the initial 
and the reduced graph we find that the diagram scales
as $l^\delta$ with 
\be
\label{pc_imp}
 \delta = \sum_k \left[ (k-4)V_k^S + (k-2-f_k)V_k^H\right]
 +E_Q +4 - 2N_C.
\ee
Here, $f_k$ denotes the number of fermion fields in  
a hard vertex, and $E_Q$ is the number of external quark 
lines. We observe that in general the scaling dimension 
$\delta$ still increases with the number of higher order 
vertices, but now there are two important exceptions. 

 First we observe that the power counting for hard vertices
is modified by a factor that counts the number of fermion lines in the
vertex. It is easy to see that four-fermion operators without extra
derivatives are leading order ($k-2-f_k=0$), but terms with more than
four fermion fields, or extra derivatives, are suppressed. This result
is familiar from the effective field theory analysis of theories with
short range interactions \cite{Shankar:1993pf,Polchinski:1992ed}.

 The second observation is that the number of fermion loops that 
become disconnected if soft gluons are removed, $N_C$, reduces the 
power $\delta$. Each disconnected loop contains at least one power 
of the coupling constant, $g$, for every soft vertex. As a result, 
fermion loop insertions in gluon $n$-point functions spoil the power 
counting if the gluon momenta satisfy $l\sim g\mu$. This implies that 
for $l<g\mu$ the high density effective theory becomes non-perturbative 
and fermion loops in gluon $n$-point functions have to be resummed. 
This resummation corresponds to the familiar hard dense loop (HDL)
resummation \cite{Braaten:1989mz,Braaten:1991gm}. Note, however, 
that in the high density effective theory we do not perform
a hard dense loop resummation of Green functions with external
fermion lines.  

 There is a simple generating functional for hard dense loops
in gluon $n$-point functions which is given by 
\cite{Braaten:1991gm}
\be 
\label{S_hdl}
{\cal L}_{HDL} = -\frac{m^2}{2}\sum_v \,G^a_{\mu \alpha} 
  \frac{v^\alpha v^\beta}{(v\cdot D)^2} 
G^b_{\mu\beta},
\ee
where $m^2=N_f g^2\mu^2/(4\pi^2)$ is the dynamical gluon mass and 
the sum over patches corresponds to an average over the direction 
of $\vec{v}$. For momenta $l<g\mu$ we have to add the HDL generating 
functional to the HDET effective action, ${\cal L}_{HDET} \to 
{\cal L}_{HDET}+{\cal L}_{HDL}$. This means that we work with 
hard dense loop resummed gluon propagators and vertices. If 
we analyze the low energy behavior in the vicinity of a generic 
point on the Fermi surface then there is no double counting involved,
because we no longer have to consider sums over patches. 
It is interesting to note that the velocity index on the 
quark field acts like a flavor label, and that the diagrams 
selected by the large $\mu$ limit are the diagrams of the 
large $N_f$ approximation. 

 The hard dense loop action describes static screening of
electric fields and dynamic screening of magnetic modes. 
Since there is no screening of static magnetic fields 
low energy gluon exchanges are dominated by magnetic 
modes. The resummed transverse gauge boson propagator is 
given by
\be
\label{d_trans}
D_{ij}(k) = \frac{\delta_{ij}-\hat{k}_i\hat{k}_j}{k_0^2-\vec{k}^2+
i\eta |k_0|/|\vec{k}|} ,
\ee
where $\eta=\frac{\pi}{2}m^2$ and we have assumed that 
$|k_0|<|\vec{k}|$. We observe that the gluon propagator 
becomes large in the regime $k\sim (\eta k_0)^{1/3}\gg k_0$.
This implies that the power counting for very low energy 
gluons has to be modified. Landau damped gluons satisfy 
the scaling laws $k^0\sim l$ and $|\vec{k}|\sim l^{1/3}$. 
As we shall see in the next section, this modified scaling 
relation has important consequences for the structure of the 
fermion self energy. 

\section{Dyson-Schwinger equation}
\label{sec_ds}

\begin{figure}
\includegraphics[width=16cm]{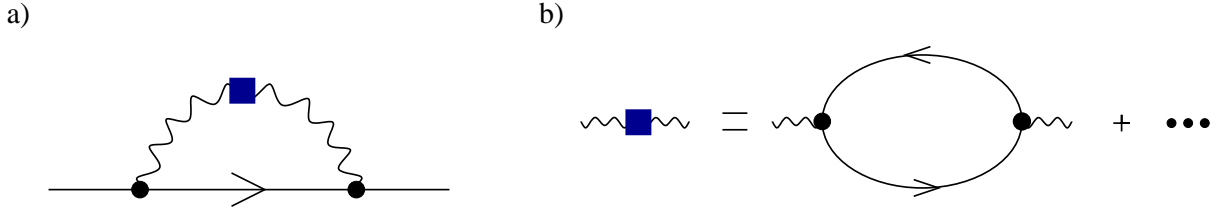}
\caption{Fig.~a) shows the dominant contribution to the
fermion self energy in the high density effective theory. 
The solid square denotes an insertion of the gluon self
energy, see Fig.~b). The dot denotes the free quark-gluon
vertex.}
\label{fig_sig1l}
\end{figure}

The leading logarithmic term in the one-loop fermion self 
energy in the high density effective theory is determined by 
the Feynman diagram shown in Fig.~\ref{fig_sig1l}. We find 
\cite{Schafer:2003jn}
\be 
\label{sig}
\Sigma(\omega,l) = \frac{g^2}{9\pi^2} \omega \log
\left(\frac{\Lambda}{\omega}\right),
\ee
where $\omega=l_0$ and $\Lambda$ is a cutoff. This result implies
that for $\omega\sim \Lambda\exp(-9\pi^2/g^2)$ the perturbative 
correction is of order 1, and higher order terms of the form 
$g^{2n}\log^n(\Lambda/\omega)$ may have to be included. In order 
to study this problem we consider the Dyson-Schwinger equation 
for the fermion self energy
\be 
\label{ds}
 -i\Sigma(p) = - \int \frac{d^4k}{(2\pi)^4}
  \Gamma_\mu^a S(p+k)\Gamma_\nu^b D^{ab}_{\mu\nu}(k).
\ee
Here, $-i\Sigma(p)=S^{-1}(p)-S_0^{-1}(p)$ is the fermion self energy, 
$D_{\mu\nu}^{ab}$ is the gluon propagator, and $\Gamma_\mu^a$ is 
the quark-gluon vertex function. Following the arguments given 
in the previous section we consider the contribution from transverse
gauge bosons only and use the HDL resummed transverse propagator. 
We will use the free quark-gluon vertex $\Gamma^a_i=gv_i\lambda^a/2$. 
We shall solve the Dyson-Schwinger equation for the quark propagator 
but we will not solve a self-consistency equation for the gluon 
propagator or the quark-gluon vertex. We will justify these 
assumptions below. 


 We note that the infrared divergence in the fermion self energy 
depends only on the energy and not on the momentum of the quark. 
Fermion momenta scale as $l_p\sim l$ while gluon momenta scale as 
$|\vec{k}|\sim l^{1/3}$. As a consequence we can neglect the 
dependence on the external fermion momentum and we will assume
that the quark self energy is a function of the energy only. The 
Dyson-Schwinger equation is  
\be
\label{sigma_0}
-i\Sigma(p_0) = g^2 C_F\int\frac{d^4k}{(2\pi)^4}
  \frac{1-(\vec{v}\cdot\hat{k})^2}
       {p_0+k_0-l_{p+k}+\Sigma(p_0+k_0)} 
  \, \frac{1}{k_0^2-\vec{k}^2+i\eta |k_0|/|\vec{k}|} ,
\ee
where $l_p=\vec{v}\cdot\vec{p}-\mu$ and $C_F=(N_c^2-1)/(2N_c)$. 
After analytic continuation to euclidean space we have 
\be 
\label{sigma_1}
\Sigma(p_4) = g^2C_F \int \frac{dk_4}{2\pi}\int 
  \frac{k^2dk}{(2\pi)^2}\int_{-1}^1 dx 
    \frac{1-x^2}{i(p_4+k_4)-l_p-k x+i\Sigma(p_4+k_4)}
   \, \frac{1}{k_4^2+k^2+\eta \frac{|k_4|}{k}},
\ee
where $l_k=\vec{v}\cdot\vec{k} \equiv  k x$ and we have 
defined $\Sigma(p_4)\equiv\Sigma_E(p_4)\equiv-i\Sigma(p_0)$. 
Because $l_p\ll |\vec{k}|$ we can neglect the dependence on 
$l_p$. The angular integration can be carried out analytically, 
leaving
\bea
\label{sigma_num}
\Sigma(p_4) &=&\frac{g^2 C_F}{4 \pi^3} 
   \int dk_4 \int k dk \left( \frac{(p_4+k_4+\Sigma (p_4+k_4))^2+k^2}{k^2} 
   \arctan \left( \frac{k}{p_4+k_4 +\Sigma (p_4+k_4)} \right) 
   \right. \nonumber \\
&&\qquad \qquad \qquad \qquad \qquad\mbox{} \left. 
   -\frac{p_4+k_4 +\Sigma (p_4+k_4)}{k} \right) \, 
    \frac{1}{k_4^2+k^2+\eta \, \frac{|k_4|}{k}} \, ,
\eea
This expression will be analyzed numerically in the next section. 
In order to derive an approximate analytic expression we note that 
the gluon propagator becomes large in the regime $k\sim (\eta k_4)^{1/3}
\gg k_4$. Therefore, we can in first approximation neglect the 
$k_4^2$ term in the gluon propagator. In addition to that, we can 
approximate $1-x^2\simeq 1$ in the numerator of equ.~(\ref{sigma_1}). 
We get
\be 
\Sigma(p_4) = 2C_Fg^2 \int \frac{dk_4}{2\pi}\int 
  \frac{kdk}{(2\pi)^2}
 \arctan\left(\frac{k}{p_4+k_4+\Sigma(p_4+k_4)}\right) 
 \frac{1}{k^2+\eta \frac{|k_4|}{k}}.
\ee
The non-analytic contribution to the self-energy can be 
extracted from 
\bea
\label{sigma_2}
\frac{d}{dp_4}\Sigma(p_4) &=&  2g^2C_F 
 \int \frac{dk_4}{2\pi}\int \frac{kdk}{(2\pi)^2}
 \Bigg\{ \pi \delta (k_4+p_4+\Sigma(p_4+k_4))  \nonumber \\
 && \mbox{}\hspace{1cm}
   -\frac{k}{(k_4+p_4+\Sigma(k_4+p_4))^2+k^2}
 \Bigg\}
\frac{1+\Sigma'(p_4+k_4)}{k^2+\eta \frac{|k_4|}{k}}.
\eea
Only the first term in the curly brackets has a logarithmic 
singularity in the limit $p_4\to 0$. We get
\be
\label{sigma_3}
\Sigma(p_4) \simeq \frac{g^2C_Fp_4}{4\pi^2} 
 \int dk \frac{k}{k^2+\eta \frac{p_4}{k}}
 \simeq  \frac{g^2C_F}{12\pi^2}p_4 \log\left( 
  \frac{\Lambda}{p_4}\right) ,
\ee
which is equal to the result of the one-loop calculation. 
This implies that as long as $g$ is small the one-loop 
self energy is a solution of the Dyson-Schwinger equation even in 
the regime $g^2\log(\Lambda/p_4)\sim 1$. It also means
that there are no contributions of the form $g^{2n}[\log(
\Lambda/p_4)]^n$ with $n>1$. A similar conclusion was
reached by Polchinski in his analysis of the spinon gauge 
theory in 2+1 dimension \cite{Polchinski:ii}. 

We finally return to the question of including vertex corrections in
the Dyson-Schwinger equation. Brown et al.~showed that vertex 
corrections are not logarithmically enhanced except in a small 
kinematic window where the collinear momentum transfer, 
$\Delta l_p$, is much smaller than the energy transfer, 
$\Delta\omega$ \cite{Brown:2000eh,Schafer:2003jn}. This 
conclusion is unchanged if self energy corrections to the 
propagator are included. The regime $\Delta l_p\ll\Delta 
\omega$ does not contribute to the Dyson-Schwinger equation
at leading order in the coupling constant. As a consequence 
vertex corrections do not have to be included at leading 
order in the weak coupling limit. This result is analogous
to Migdal's theorem for the electron-phonon interaction
\cite{Migdal:1958}. We should emphasize, however, that the 
physical picture that underlies Migdal's theorem is quite
different from what happens in the QCD case. 

 We have also checked that fermion self energy insertions 
do not modify the gluon self energy, Fig.~\ref{fig_sig1l}b, 
at leading order in the coupling constant. This result is 
related to the fact that the leading term in the self energy 
depends only on the energy, $\omega$, and not on the momentum, 
$l_p$, of the quark.

\section{Numerical Analysis}
\label{sec_num}

\begin{table}
  \begin{center}
    \begin{tabular}{c|c|c|c|c}
     \, $\mu$ [GeV] \, &  \hspace{0.3cm} $\alpha_s$\hspace{0.3cm} 
     & \, $m$ [GeV] \, 
     & \, $\omega_{bcs}$ [GeV] \,  
     & \, $\omega_{\it nfl}$ [GeV] \, \\
     \hline
       0.5 & 1.1   & 0.50 & $2.2 \cdot 10^{-2}$ & $7 \cdot 10^{-4}$   \\
       1   & 0.52  & 0.70 & $2.2 \cdot 10^{-2}$ & $8 \cdot 10^{-7}$   \\
       100 & 0.12  & 33   & $1.1 \cdot 10^{-2}$ & $2 \cdot 10^{-25}$  \\
 $10^{10}$ & 0.029 &$1.7 \cdot 10^9$ & $7.9 \cdot 10^{-1}$ 
           & $ 1 \cdot 10^{-98}$
    \end{tabular}
\vspace{0.2cm}
\caption{
Characteristic scales in dense quark matter. We compare 
the dynamical screening mass $m$, the BCS scale $\omega_{bcs}$,
and the scale of non-Fermi liquid effects $\omega_{\it nfl}$. 
We also show the one-loop running coupling constant $\alpha_s$ 
evaluated at the chemical potential $\mu$.}
\label{tab_scales}
\end{center}
\end{table} 

 In the previous section we presented analytic arguments
which suggest that the infrared enhancement in the fermion
self energy is one-loop exact in the weak coupling limit. 
In this section we shall strengthen these arguments by 
performing a numerical study of the Dyson-Schwinger 
equation (\ref{sigma_num}). This will also provide an 
estimate of the size of higher order corrections at 
non-asymptotic densities. 

\begin{figure}
\includegraphics[width=14cm]{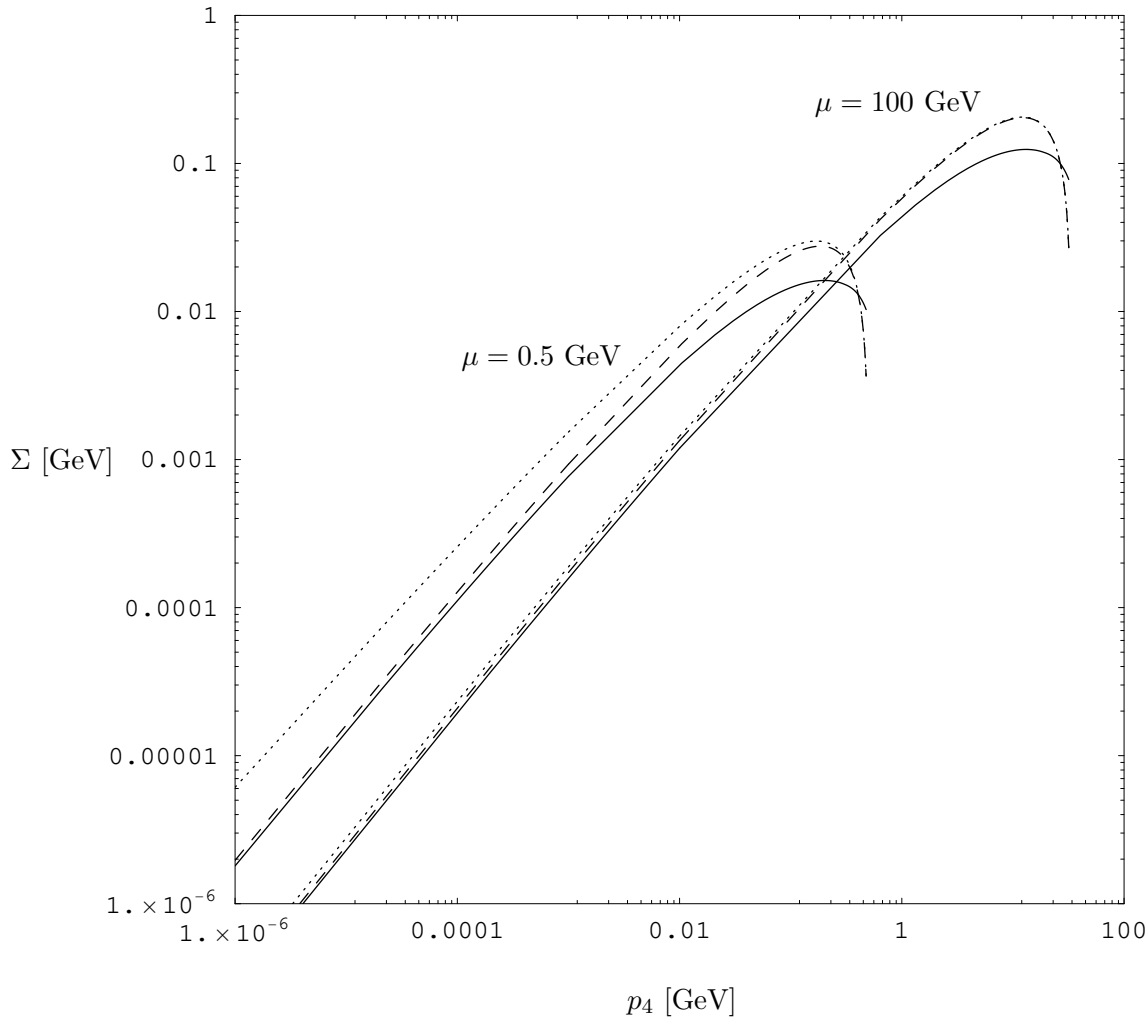}
\flushleft \vspace*{-1.0cm} \hspace*{8.1cm}  $p_4$ [GeV]
\flushleft \vspace*{-8.3cm} \hspace{-0.1cm} $\Sigma$ [GeV]
\flushleft \vspace*{-2.5cm} \hspace{5.9cm} $\mu = 0.5$ GeV
\flushleft \vspace*{-4.5cm} \hspace{10.6cm} $\mu=100$ GeV
\vspace*{11.8cm}
\caption{Numerical solution of the Dyson-Schwinger equation 
for the fermion self energy. We show the euclidean self 
energy $\Sigma(p_4)$ as a function of $p_4$. The two sets
of curves correspond to two different chemical potentials, 
$\mu =0.5$ GeV and $\mu=100$ GeV. The solid lines show the 
numerical solution, the dashed curves show the analytic result 
for the leading logarithm, and the dotted curves show a
power-like self energy $\Sigma(p_4)\sim p_4(\Lambda/p_4)^\gamma$.}
\label{fig_selfenergy}
\end{figure}

  Let us first give numerical estimates for the relevant 
scales. We shall assume that the strong coupling constant
$\alpha_s$ is given by the $N_f=3$ one-loop running coupling 
constant evaluated at the scale $\mu$. In  Table \ref{tab_scales} 
we compare the dynamical screening scale $m$, the scale of 
superconductivity $\omega_{bcs}\sim b_0\mu g^{-5} \exp(-
c_{bcs}/g)$, and the scale of non-Fermi liquid effects 
$\omega_{\it nfl} \sim m \,\exp(-c_{\it nfl}/g^2)$. We use 
$c_{bcs}=3\pi^2/\sqrt{2}$ and $c_{\it nfl}=9\pi^2$ given 
above as well as $b_0=512\pi^4 \exp(-(\pi^2+4)/8)$.
Even if the chemical potential is very large the screening 
scale is close to the Fermi energy. The scale of superfluidity 
varies very little whereas the scale of non-Fermi liquid effects 
becomes extremely small if the chemical is large.

 We have solved the Dyson-Schwinger equation for two different
values of the chemical potential, a low value of $\mu=0.5$ GeV 
relevant for the physics of neutron stars and an asymptotically 
large value of $\mu=100$ GeV. At the high scale perturbation 
theory is expected to be applicable but at the lower scale 
the coupling is not small and the usefulness of perturbation
theory is in doubt. It was shown, however, that the numerical
solution of the gap equation in the superconducting phase is
close to the asymptotic solution even if the coupling is not
small, and that the gap on the Fermi surface is quite consistent
with values obtained from phenomenological models \cite{Schafer:1999jg}. 
We have chosen the cutoff of the effective theory to be equal to 
the dynamical screening scale $m$. 

\begin{figure}
\includegraphics[width=14cm]{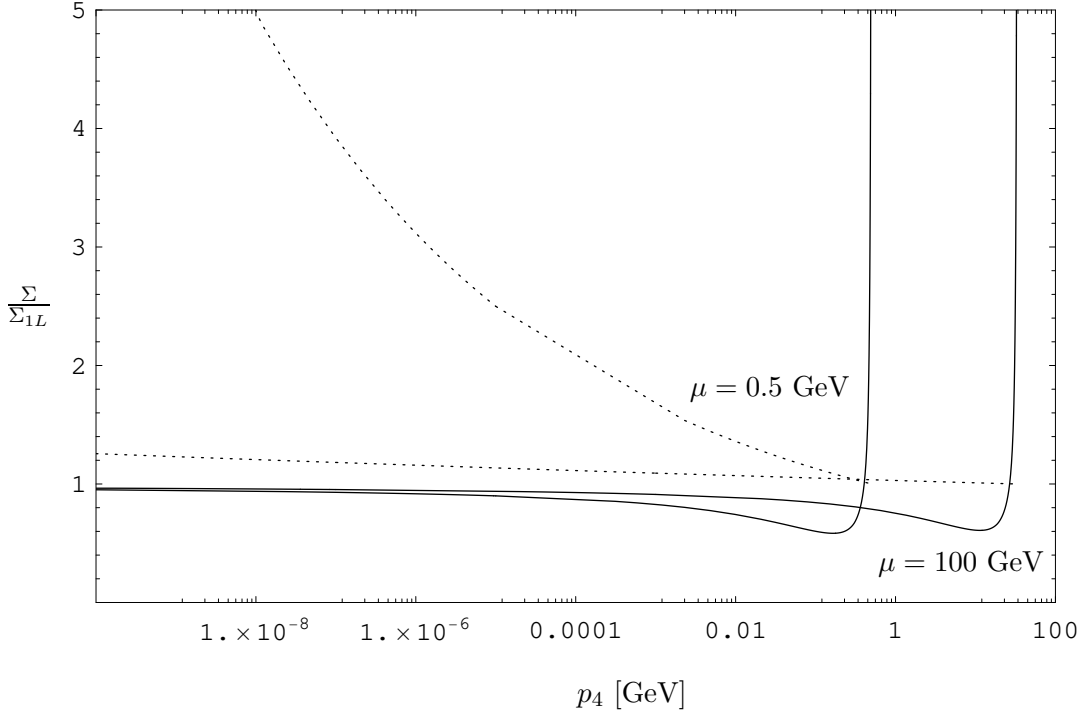}
\flushleft \vspace*{-0.4cm} \hspace*{8.1cm}  $p_4$ [GeV]
\flushleft \vspace*{-6.3cm} \hspace{0.5cm} $\frac{\Sigma}{\Sigma_{1L}}$ 
\flushleft \hspace*{9.6cm} $\mu = 0.5$ GeV
\flushleft \vspace*{ 1.2cm} \hspace*{12.1cm} $\mu = 100$ GeV
\vspace*{1.7cm}
\caption{Comparison between the numerical solution of the 
Dyson-Schwinger equation and the analytic leading log result. 
The solid lines show the ratio of the numerical solution over the 
leading log term for two different chemical potentials. The dotted 
lines show the same ratio for the power-like self energy $\Sigma
(p_4)\sim p_4(\Lambda/p_4)^\gamma$.}
\label{fig_ratios}
\end{figure}

  Our results are shown in Fig.~\ref{fig_selfenergy}. The solid 
line shows the numerical solution and the dashed line shows the 
analytic result for the leading logarithm, equ.~(\ref{sig}).
For comparison we also show a self energy function that scales 
as a fractional power of energy, $\Sigma(p_4)=p_4(\Lambda/p_4)^\gamma$ 
with $\gamma=g^2/(9\pi^2)$. This behavior was proposed by Boyanovsky 
and de Vega on the basis of a renormalization group study
\cite{Boyanovsky:2000bc}. Since
\be
\label{expansion}
p_4 \, \left( \frac{p_4}{\Lambda} \right)^\gamma \simeq 
   p_4 \left\{ 1  + \gamma \log \left( \frac{p_4}{\Lambda} \right) 
 + O \left( \gamma^2 \log^2 \left( \frac{p_4}{\Lambda} 
   \right)\right) \right\}
\ee
this functional form starts to deviate substantially from 
the one-loop result for energies below the non-Fermi liquid
scale, see Table \ref{tab_scales}.

 For very small values of the energy, $p_4\ll \Lambda$, we
find excellent agreement between the numerical results and 
the leading log expression. This is seen even more clearly in
Fig.~\ref{fig_ratios} where we show the ratio of the numerical
solution over the leading logarithm. Significant deviations only 
occur for energies near the cutoff, but some higher order corrections 
are still present for energies one or two orders of magnitude 
below the cutoff. Our results show no evidence for higher order 
terms of the form $g^{2n}\log^n(p_4)$ with  $n>1$ and support 
the arguments given in the previous section. We conclude that 
in the limit of weak coupling and low energy the self energy 
is given by equ.~(\ref{sig}) even in the regime where the 
logarithm is large.

\section{Renormalization group}
\label{sec_rg}

 Boyanovsky and Vega argued, on the basis of a renormalization
group analysis, that the inverse quark propagator is of the 
form $S^{-1}(\omega,l)=\omega(\omega/\Lambda)^\gamma-l$ where
$\gamma=g^2/(9\pi^2)$ in the weak coupling limit. A similar 
result in the QED case was presented by Gan and Wong \cite{Gan:1993}.
These arguments contradict the solution of the Dyson-Schwinger 
equation presented in the preceding two sections. In order to clarify
the situation we consider the renormalization group equation 
for the two-point function in the high density effective theory. 
We shall consider the leading order lagrangian
\be 
{\cal L} = \psi^\dagger_v \left(\omega-v_F l\right)\psi_v
 + gv_F\psi^\dagger_v\hat{v}\cdot\vec{A} \psi_v + \ldots ,
\ee 
where we have explicitly included the Fermi velocity $v_F$.
This is necessary because at finite density Lorentz invariance
is broken and we need separate wave function renormalization 
factors for the energy and momentum dependent terms in the 
action. The one-loop fermion self energy is given by 
\be 
\label{sig_1lv}
 \Sigma(\omega,l)= \frac{g^2v_F}{9\pi^2}\omega 
  \log\left(\frac{\Lambda}{\omega}\right) ,
\ee
where we have neglected terms that do not contain logarithms. 
Equ.~(\ref{sig_1lv}) has a logarithmic divergence in the 
effective field theory. This divergence can be removed by 
adding a counter-term to the lagrangian. We define the 
bare lagrangian ${\cal L}_0={\cal L}+{\cal L}_{ct}$
\bea
{\cal L}_0 &=& \psi^\dagger_v \left(Z\omega- ZZ_F v_F l\right)\psi_v
 + Z_g gv_F\psi^\dagger_v\hat{v}\cdot\vec{A} \psi_v  \nonumber \\
 &=& \psi^\dagger_{0,v} \left(\omega-v_{0,F} l\right)\psi_{0,v}
 + g_0v_{0,F}\psi^\dagger_{0,v}\hat{v}\cdot\vec{A} \psi_{0,v}  ,
\eea
as well as the bare fields and coupling constants
\be
\psi_{0,v} = Z^{1/2}\psi_v,  \hspace{1cm} 
v_{0,F}    = Z_F v_{F},    \hspace{1cm}
 g_0       = \frac{Z_g}{ZZ_F}g.
\ee
Equ.~(\ref{sig_1lv}) implies that at one-loop order $Z\sim \log
(\Lambda)$ and that $ZZ_F=1$. We showed in the previous section
that, in the kinematic regime of interest, there is no logarithmic 
divergence in the quark-gluon vertex. As a consequence we can take
$Z_g=1$. Equ.~(\ref{sig_1lv}) also suggests, however, that the 
effective coupling constant is
\be 
\alpha = \frac{g^2v_F}{4\pi},
\ee
which acquires an anomalous dimension because of the scaling 
of the Fermi velocity. The bare and renormalized Green functions 
are related by 
\be 
G^{(n)}_0(\omega_i,v_{0,F}l_i,\alpha_0) = Z^{n/2} 
 G^{(n)}(\omega_i,v_{F}l_i,\alpha),
\ee
where $n$ denotes the number of external fermion fields, 
$(\omega_i,l_i)$ are the external energies and momenta.
Differentiating this relation with respect to $\Lambda$
gives the renormalization group equation
\be
\label{rg1}
\left\{ \Lambda\frac{\partial}{\partial \Lambda}
 + \beta(\alpha) \frac{\partial}{\partial \alpha}
 - \gamma_F(\alpha) l_i\frac{\partial}{\partial l_i}
 + \frac{n}{2} \gamma(\alpha) \right\} G^{(n)}(\omega_i,l_i,\alpha) 
 = 0 ,
\ee
where we have defined the beta function and the anomalous
dimensions
\be
\beta(\alpha) = \Lambda\frac{\partial \alpha}{\partial \Lambda},
\hspace{1cm}
\gamma(\alpha)= \Lambda\frac{\partial \log Z}{\partial \Lambda},
\hspace{1cm}
\gamma_F(\alpha)= \Lambda\frac{\partial \log Z_F}{\partial \Lambda}.
\ee
In deriving equ.~(\ref{rg1}) we have used the fact that $G^{(n)}$ depends 
on the Fermi velocity only through $\alpha$ and $v_Fl_i$. At one-loop 
order we have
\be 
\label{gam_pert}
\beta(\alpha) = -\gamma_F(\alpha)\alpha ,
\hspace{1cm}
\gamma(\alpha)= -\gamma_F(\alpha) = \frac{4\alpha}{9\pi}.
\ee
We note that the beta function is positive and the effective 
theory is infrared free. This means that the perturbative analysis
of the low energy behavior is reliable. The fact that the effective 
coupling is weak at low energy is related to the fact that the Fermi 
velocity vanishes as the quasi-particle energy goes to zero. We
have checked that the one-loop anomalous dimensions do not depend
on the gauge parameter in a generalized Coulomb gauge. In general,
of course, there is no reason to expect the anomalous dimensions
to be gauge invariant. Only physical properties of the solutions
of the renormalization group equation, such as the quasi-particle 
properties, are gauge invariant.

 Boyanovsky and de Vega solved the renormalization group equation 
under the assumption that the beta function vanishes \cite{Boyanovsky:2000bc}. 
In this case we have 
\be
\label{rg2}
\left\{ \Lambda\frac{\partial}{\partial \Lambda}
 + \gamma \left[ l\frac{\partial}{\partial l}
 + 1 \right] \right\} S(w,l,\alpha) = 0 ,
\ee
where we have used $\gamma_F=-\gamma$. We observe that the 
inverse propagator satisfies the renormalization group equation
$\{\Lambda \partial/(\partial \Lambda) +\gamma[l\partial /
(\partial l) -1]\}S^{-1}(\omega,l,\alpha)=0$. It is easy to 
see that this equation is solved by 
\be
\label{rg1_sol}
S^{-1}(\omega,l) = \omega\left( \frac{\Lambda}{\omega}
 \right)^\gamma - v_F l ,
\ee 
where we have imposed the boundary condition $S^{-1}(\omega
\!=\!\Lambda,l)=\omega-v_Fl$. Equ.~(\ref{rg1_sol}) is the result 
of Boyanovsky and Vega. We showed, however, that the beta 
function does not vanish. The complete renormalization group 
equation is 
\be
\label{rg3}
\left\{ \Lambda\frac{\partial}{\partial \Lambda}
 + \gamma \left[ \alpha\frac{\partial}{\partial\alpha}
 + l\frac{\partial}{\partial l} + 1 \right] \right\} 
           S(w,l,\alpha) = 0 .
\ee
The propagator with the one-loop self energy included
\be 
\label{rg3_sol}
S^{-1}(\omega,l) = \omega\left( 1+\gamma\log\left(
 \frac{\Lambda}{\omega}\right)\right) - v_F l 
\ee 
is a solution of the complete renormalization group equation. We 
can also study the possible presence of higher order terms
of the form $S^{-1}\sim \alpha^n\log^n(\Lambda/\omega)$.
Consider the ansatz
\be
\label{rg_ans}
S^{-1}(\omega,l)=\sum_k a_k \alpha^k \omega \left[
 \log\left(\frac{\Lambda}{\omega}\right)\right]^k -v_Fl.
\ee
Inserting this ansatz into the renormalization group equation
we obtain $a_{k+1}=a_k \gamma_1 (k-1)/(k+1)$, where we have used
the one-loop anomalous dimension $\gamma(\alpha)= \gamma_1\alpha$. 
This shows that $a_k=0$ for $k>2$ and terms of order $\alpha^2
\log^2(\Lambda/\omega)$ or higher are absent.

\section{Summary and Discussion}
\label{sec_sum}

 We have studied non-Fermi liquid effects due to unscreened
transverse gauge boson exchanges in the normal phase of high 
density QCD. We find that if the coupling is weak the fermion
self energy is given by $\gamma\omega\log(\Lambda/\omega)$ with 
$\gamma=g^2/(9\pi^2)$. This result is reliable even if $\gamma
\log(\Lambda/\omega)\sim 1$. We established this result using
two different methods, the Dyson-Schwinger equation and the 
renormalization group. In the context of the Dyson-Schwinger 
equation the absence of higher order corrections is a 
consequence of the special kinematics of ungapped fermions
interacting with Landau damped gluons. In the kinematic
regime of interest the right hand side of the Dyson-Schwinger
equation is independent of the fermion self energy. As a 
consequence, there is no difference between the one-loop
result and the self-consistent solution. In the context 
of the renormalization group the absence of higher order 
terms follows from the relations $\gamma=-\gamma_F$ and
$\beta=-\gamma_F\alpha$ \cite{Chakravarty:1995}. The relation
between the anomalous dimension of the fermion field and the Fermi 
velocity is again due to the special kinematics. The relation
between the beta function and the anomalous dimension 
of the Fermi velocity is a consequence of gauge invariance. 

 The weak coupling result implies that the quark propagator 
has a cut rather than a pole, and that the naive quasi-particle 
description breaks down. The spectral density is given by
\be
\rho(\omega) = \frac{\gamma\omega}
  {\left[\omega(1+\gamma\log(\Lambda/\omega))-l\right]^2
   +\pi^2\gamma^2\omega^2}.
\ee
For non-zero momentum $l$ this is approximately a Breit-Wigner 
distribution, but the wave function normalization and 
Fermi velocity vanish as $l\to 0$. An important consequence
of the breakdown of Fermi liquid theory is an anomalous
term in the specific heat. Ipp et al.~showed that 
\cite{Ipp:2003cj}
\be 
\label{cv_anom}
C_v^{anom}=\gamma 
    C_v^{free}\log\left(\frac{\Lambda}{T}\right)
  = N_f(N_c^2-1)\frac{g^2\mu^2 T}{72\pi^2}
  \log\left(\frac{\Lambda}{T}\right),
\ee
where $C_v^{free}=N_cN_f \mu^2 T/3$. They also computed the 
argument of the logarithm as well as terms that include fractional 
powers $T^{5/3}$ and $T^{7/3}$. Our results suggest that 
equ.~(\ref{cv_anom}) is reliable even if $g^2\log(\Lambda/T)\gg 1$. 
The anomalous term in the fermion self energy does not lead to 
an anomalous term in the thermodynamic potential $\Omega$ at 
$T=0$. The two-loop contribution to $\Omega$ is infrared finite. 
Instead, this graph has an ultraviolet divergence in the effective 
field theory. This means that the thermodynamic potential has to 
be determined in the microscopic theory. The result is 
\cite{Freedman:1976ub}
\be
\Omega = -\frac{N_fN_c\mu^4}{12\pi^2}
 \left\{ 1 - \frac{3(N_c^2-1)}{4N_c} \left(\frac{\alpha_s}{\pi}
   \right) + O(\alpha_s^2)\right\}.
\ee
In the superconducting phase the infrared enhancement in the
fermion self energy is cutoff for energies less than the gap. 
Because $\Delta\sim \exp(-c_{bcs}/g)$ the correction to the self 
energy never exceeds $\gamma\log(\Delta)\sim O(g)$. As a consequence 
non-Fermi liquid effects in the normal phase do not qualitatively 
alter the superconducting phase of QCD, but they give a correction
to the gap which is enhanced by one power of $1/g$ relative to its 
naive order in the coupling constant. This correction reduces the 
gap by a factor $\exp[-(\pi^2+4)(N_c-1)/16]\sim 0.18$, where we 
have set $N_c=3$ 
\cite{Brown:1999aq,Brown:1999yd,Wang:2001aq,Schafer:2003jn}.

 In this paper we did not study Green functions with more than 
two external fermion lines. It is not clear whether more complicated 
$n$-point functions exhibit additional infrared divergences. It would 
be interesting, for example, to study the propagation of zero sound 
in the normal phase of dense quark matter. With regard to the physics 
of neutron stars it would also be interesting to study the thermal
conductivity as well as the neutrino emissivity and opacity.

Acknowledgments: We would like to thank C.~Manuel and A.~Rebhan
for useful discussions. This work was supported in part by
US DOE grant DE-FG-88ER40388. This manuscript was completed
at the Institute for Nuclear Theory during the workshop 
on ``QCD and Dense Matter: From Lattices to Stars''. We 
thank the INT for hospitality. 


\end{document}